4# Transmission of High-Definition Video Signals Underwater using Surface Electromagnetic Waves

Igor I. Smolyaninov, Quirino Balzano, and Mark Barry

*Abstract*—A portable radio communication system operating in the 30 MHz band and capable of transmitting high-definition live underwater video images is presented. The system operation is based on launching electromagnetic surface waves propagating along water-air interface using specially designed surface wave antennas. Since the propagation length of the surface electromagnetic waves far exceeds the skin depth of bulk radio waves at the same frequency, this technique is useful for video communication underwater over distances of several meters. Also, this system appears to be efficient at communicating through the water-air interface.

## I. INTRODUCTION

WIRELESS video communication underwater remains an unsolved problem which considerably limits undersea exploration. For safety and efficiency, divers, unmanned underwater vehicles (UUV), and other subsea equipment may need to communicate with each other, or with their surface or shore-based support teams. Currently developed underwater acoustic and optical communication systems provide scarce capabilities. The limited bandwidth and large latency of acoustic communication devices does not permit efficient transmission of live high-definition video signals underwater [1,2]. On the other hand, the range of optical wireless communication systems in turbid water may be as short as ~ 0.1 m [3,4].

Very recently it was demonstrated that surface electromagnetic wave-based radio communication underwater may become a viable alternative to acoustic and optical wireless communication techniques [5,6]. Since propagation length of the surface electromagnetic waves (SEW) far exceeds the skin depth of the conventional radio waves at the same frequency [7,8], this technique may become useful for broadband radio communication underwater over practical distances. In this paper we will demonstrate that the bandwidth of such radio communication systems is large enough to achieve live high-definition video communication underwater. Moreover, we will also show that such surface wave systems may be used to directly communicate through the water-air interface, which is known to be extremely difficult for acoustic and optical communication systems [9].

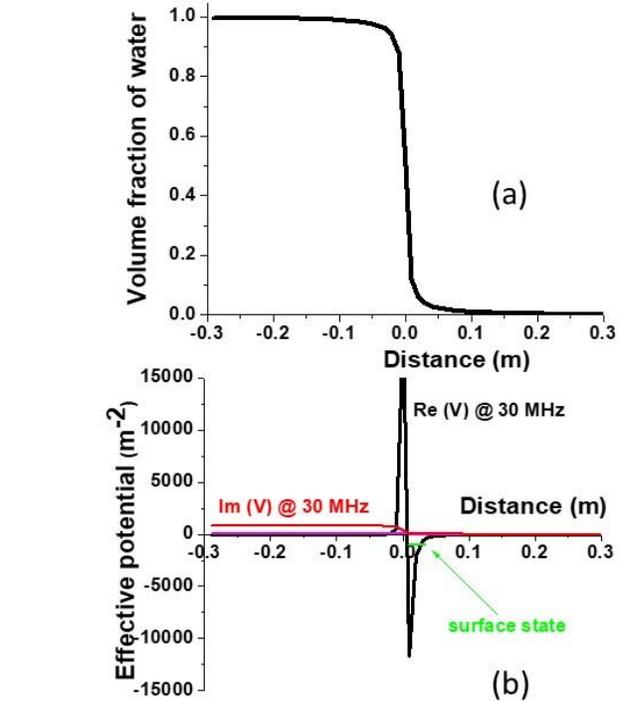

**Fig. 1.** (a) Assumed model of the ~ 5 cm wavy transition layer near the air-water interface. (b) Real (black) and imaginary (magenta and red) parts of the effective potential energy near the air-water interface calculated for the 30 MHz frequency band using Eq.(1). The modeled water salinity was 0.2% (magenta) and 3.5% (red). The SEW state is indicated by the green arrow.

## II. THEORETICAL BACKGROUND

Let us consider the macroscopic Maxwell equations in a non-magnetic ($\mu_r=1$) medium, in which the dielectric permittivity $\varepsilon(z)$ is continuous, and it depends only on z coordinate (as shown in Fig.1a). The continuous function $\varepsilon(z)$ describes a transition from water to air which occurs gradually around $z=0$, due to the waviness of the water surface. In such a geometry we may search for source-free modes propagating in the *x* direction as $e^{i(kx-\omega t)}$ where *k* is the wave vector. As described in detail in ref. [7], surface electromagnetic wave solutions of the

I. I. Smolyaninov and M. Barry are with the Saltenna LLC, McLean, VA 22102-4903 USA (e-mail: igor.smolyaninov@saltenna.com; mark.barry@saltenna.com).

Q. Balzano is with the Department of Electrical and Computer Engineering, University of Maryland, College Park, MD 20742 USA (e-mail: qbalzano@umd.edu).

macroscopic Maxwell equations may be found only for the TM ($E_z \neq 0$) polarization as solutions of the effective Schrodinger equation

$$-\frac{\partial^2 \psi}{\partial z^2} + \left(-\frac{\varepsilon(z)\omega^2}{c^2} - \frac{1}{2}\frac{\partial^2 \varepsilon}{\varepsilon \partial z^2} + \frac{3}{4}\frac{(\partial \varepsilon/\partial z)^2}{\varepsilon^2}\right)\psi = -\frac{\partial^2 \psi}{\partial z^2} + V\psi = -k^2 \psi \quad (1)$$

where the effective wave function $\psi$ is introduced as $E_z = \psi/\varepsilon^{1/2}$. In this wave equation $-k^2$ plays the role of the effective energy, and $V(z)$ plays the role of the effective potential. Note that when the gradient terms is Eq.(1) dominate, $V(z)$ may remain almost real even if $\varepsilon(z)$ itself is imaginary [4].

As an example, the real and imaginary parts of the complete effective potential energy defined by Eq.(1) have been calculated at 30 MHz and plotted in Fig.1b for a wavy interface between air and water (0.2% pool water salinity and 3.5% seawater salinity has been assumed, while the dielectric properties of salt water were taken from [10]). Note that the real part of the effective potential is rather insensitive to water salinity. Such a potential well will always have at least one bound state, which gives rise to a SEW solution having an almost pure real longitudinal wave vector $k$. This surface mode has a long propagation length $L$ along the air-water boundary [7,8]. By treating Im($V$) as a perturbation, it may be estimated as $L=\text{Im}(k)^{-1}$, where

$$\text{Im}(k) \sim \frac{1}{2}\int \text{Im}(V)dz \quad (2)$$

and the integration is performed over the classically accessible region [11]. As described in detail below, this surface mode has been used in transmission of high-definition video signals underwater.

### III. EXPERIMENTAL CONFIGURATION

The block diagram of our software-defined radio (SDR)-based portable underwater video communication system operating in the 30 MHz band is shown in Fig.2. The electronic and control units of the system have been provided by Software Defined Radio Solutions LLC. This system is able to transmit and receive high-definition video signals at a center frequency of 30 MHz, with up to 107 dB of attenuation between transmitter and receiver.

The transmitter consists of a video camera with High-Definition Multimedia Interface (HDMI) output, an Advanced Television System Committee (ATSC) digital TV modulator, a Spyverter frequency converter (used to convert the signal from 150 MHz to 30 MHz), and two Nooelec Lana low noise amplifiers (LNA). The maximum transmit power was a very modest 20 mW (13dBm). The receiver consists of two Nooelec Lana LNAs, a Spyverter frequency converter (used to convert the signal from 30 MHz to 150 MHz), an Ettus Research USRP B200mini SDR/Cognitive Radio and Intel NUC Frost Canyon Slim computer (used to convert the signal from 150 MHz to a standard TV channel at 79 MHz), and an ATSC digital TV demodulator. These transmitter and receiver units were submerged underwater and connected to their respective SEW underwater antennas operating in the 30 MHz band. The design of similar 50 MHz SEW antennas is described in detail in [5,6]. The antenna operation is made possible due to implementation of an impedance matching enclosure, which is filled with de-ionized water. Enhanced coupling to surface electromagnetic waves is enabled by the enhancement of the electromagnetic field at the antenna apex. These features allow us to make antenna dimensions considerably smaller compared to typical free space designs. They also considerably improve the coupling of electromagnetic energy to the surrounding water.

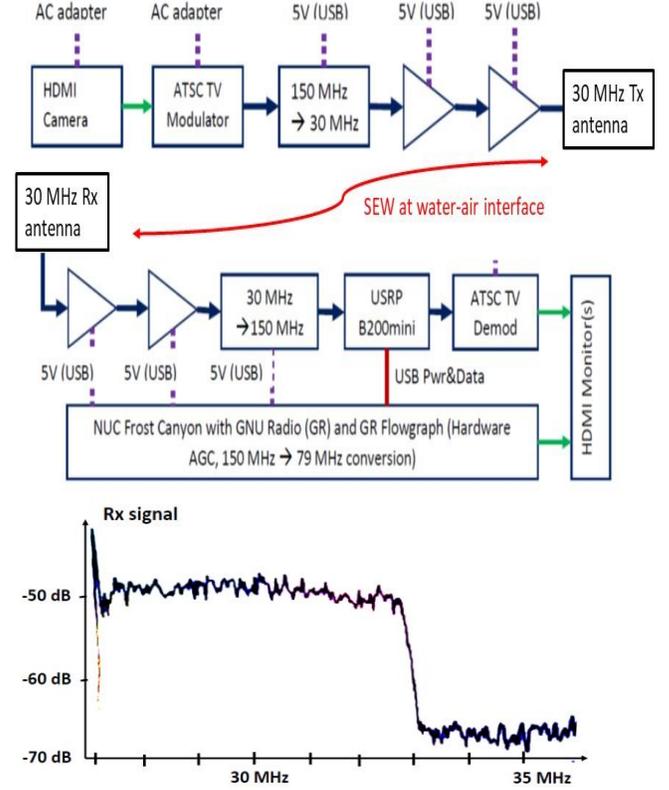

**Fig. 2.** Block diagram of the portable underwater video communication system operating in the 30 MHz band and the spectrum of the video signal received underwater. The video signal bandwidth is 6 MHz.

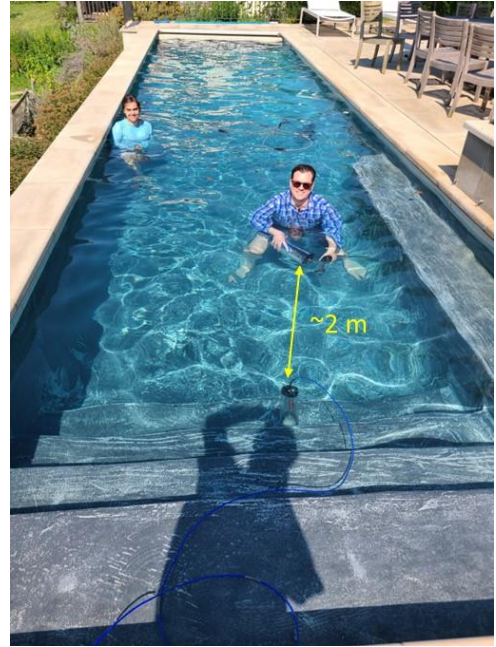

**Fig. 3.** Experimental geometry of the pool testing. The 30 MHz video signal is transmitted using an underwater SEW antenna using 20 mW (13dBm) transmit power. The measured water salinity in the pool was 0.2%, which corresponds to 20 cm skin depth at 30 MHz.

The typical spectrum of live high-definition video signal received underwater is shown in Fig.2. Performance of this video communication system has been extensively tested in a well-controlled swimming pool environment, as illustrated in Fig.3.

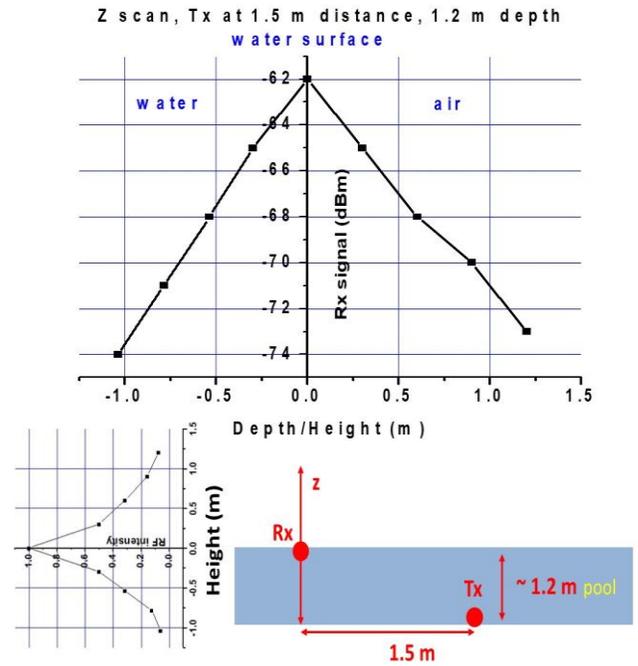

**Fig. 5.** Typical Rx signal dependence as a function of z measured across the air-water interface in the experimental geometry shown on the bottom. The inset on the left shows the same measured Rx signal in linear scale. The measured field profile matches the anticipated SEW signal.

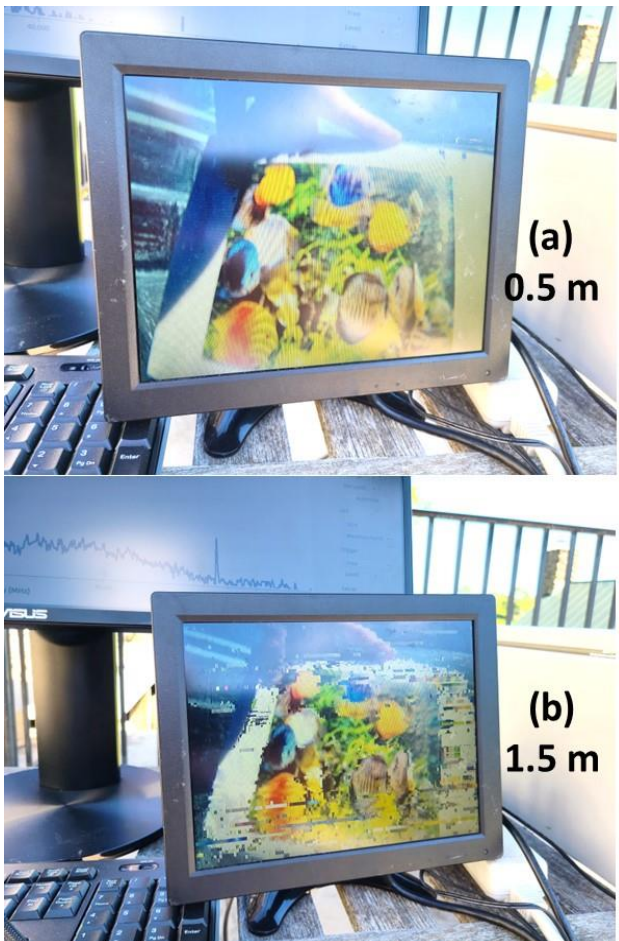

**Fig. 4.** Deterioration of high-definition video signal as a function of underwater distance (indicated in the respective photos). The Tx and Rx depth underwater is 0.54 m.

## IV. RESULTS AND DISCUSSION

As expected, we were able to transmit and receive live high-definition video images underwater, as illustrated in Fig.4. The received SEW signal and its quality was studied as a function of depth and distance underwater. The typical Rx signal intensity as a function of $z$ measured across the air-water interface is shown in Fig.5. The same antenna was used below and above the interface. As expected, the measured field profile matched the SEW distribution, which falls off exponentially away from the water-air interface into both media [7,8]. Note that this experiment also proves that the SEW transmission system is quite efficient at communicating through the water-air interface, which is known to be extremely difficult for acoustical and optical wireless communication systems [9].

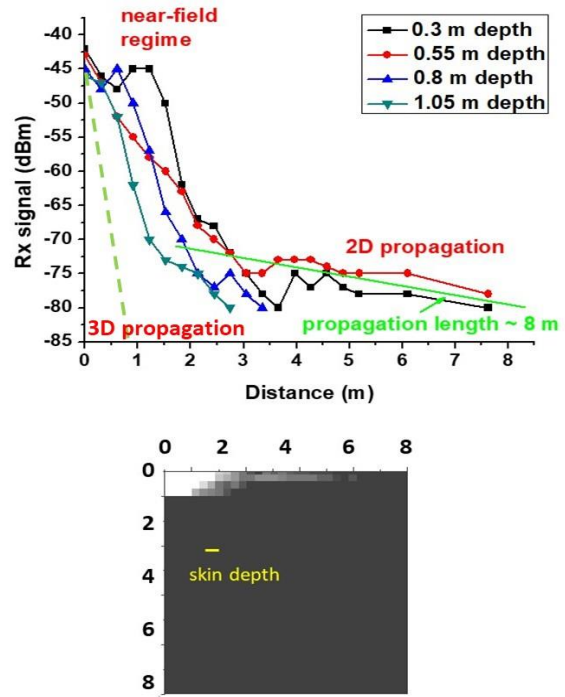

**Fig. 6.** Rx signal measured as a function of distance between the Tx and Rx. During each series the Tx and Rx depth (indicated in the box next to the series color code) was kept the same. The inset shows the same data plotted as a 2D halftone map. These plots confirm the SEW transmission mechanism underwater.

The detailed measurements of the received signal as a function of distance between the Tx and Rx at different depth underwater are presented in Fig.6. During each series the Tx and Rx depth (indicated in the box next to the series color code) was kept the same. These data clearly show the transition from the near-field regime near the antenna to 2D signal propagation in the far field. Note that the measured $L$=8 m propagation length of the SEW signal far exceeds the classical $\delta$=20 cm skin depth at $\nu$=30 MHz in 0.2% salinity pool water defined as [12]

$$\delta = \sqrt{\frac{1}{\pi\mu_0\sigma\nu}}, \qquad (3)$$

where $\sigma$ is water conductivity (compare the continuous green line corresponding to 2D propagation with the dashed green line indicating 3D propagation characterized by the 20 cm skin depth). These results are consistent with theoretical expectations [7,8] and our earlier SEW transmission experiments performed underwater at 50 MHz [5,6].

Finally, the performance of our system has been tested in seawater. As designed and built, the system operates at a maximum transmit power of 20 mW, and it was capable of maintaining video link over distances of the order of 0.5 m in seawater. It was verified that system performance was unaffected by water turbidity. We expect that further increase in transmit power will lead to further improvement of its distance performance in seawater.

Indeed, based on the measured $L$=8 m propagation length in water salinity $n$=0.2% pool water (see Fig. 6), and on the theoretical predictions based on Eqs.(1,2), the theoretically expected 2D propagation length $L_{SW}$ in seawater (water salinity $n_{SW}$=3.5%) at 30 MHz is supposed to reach $L_{sw}=nL/n_{sw}$~0.46 m, which is consistent with our experimental observations. Therefore, by increasing the transmit power to 10 W, we may expect the communication range to increase to at least 8 m in seawater (assuming similar ~ -110 dBm Rx sensitivity). We also anticipate that the described SEW antennas may be made multi-spectral, so that the communication bandwidth at a given distance/depth combination may be optimized under SDR control, shifting to larger bandwidths over shorter distances.

## V. CONCLUSION

In conclusion, we have demonstrated an underwater portable software-defined radio communication system operating in the 30 MHz band and capable of transmitting high-definition live video images underwater. The system operation is based on launching surface electromagnetic waves propagating along the water-air interface using specially designed surface wave antennas [5,6]. Since the propagation length of the surface electromagnetic waves far exceeds the classical skin depth of conventional radio waves at the same frequency, this technique is useful for video communication underwater over several meters distances. The described system also appears to be efficient at communicating through the water-air interface. We anticipate that given the 6 MHz bandwidth of our video communication system, its operation may be further shifted down towards carrier frequencies in the 10-15 MHz range, which together with increased power budgets would further increase communication distance and depth underwater.

Our SEW-based underwater communication scheme may find applications in frogman-to-frogman communication, UUV swarming and mesh networked UUVs, offshore oil platforms, etc. We should also note that the demonstrated propagation of novel surface electromagnetic waves along the water-air interface may lead to the development of novel underwater imaging techniques, which would be sensitive to dielectric properties of underwater objects. Such a novel capability would differentiate the SEW-based underwater imaging from the commonly used acoustic imaging techniques [13].

ACKNOWLEDGMENT

We would like to thank V. I. Smolyaninova and D. Young for experimental help.